\documentclass[unnumsec,webpdf,contemporary,large]{oup-authoring-template}%

\graphicspath{{Fig/}}

\theoremstyle{thmstyleone}%
\theoremstyle{thmstyletwo}%
\theoremstyle{thmstylethree}%

\begin{document}

\journaltitle{Integrative and Comparative Biology}
\DOI{DOI HERE}
\copyrightyear{2022}
\pubyear{2022}
\access{Advance Access Publication Date: Day Month Year}
\appnotes{Symposium}

\firstpage{1}


\title[Oxygenation-Controlled Collective Dynamics in Aquatic Worm Blobs]{Oxygenation-Controlled Collective Dynamics in Aquatic Worm Blobs}

\author[1]{Harry Tuazon}
\author[1]{Emily Kaufman}
\author[2]{Daniel I. Goldman}
\author[1,$\ast$]{M. Saad Bhamla}

\authormark{Tuazon, H. et al.}

\address[1]{\orgdiv{School of Chemical and Biomolecular Engineering}, \orgname{Georgia Institute of Technology}, \orgaddress{\postcode{Atlanta}, \state{Georgia}, \country{United States}}}
\address[2]{\orgdiv{School of Physics}, \orgname{Georgia Institute of Technology}, \orgaddress{\postcode{Atlanta}, \state{Georgia}, \country{United States}}}

\corresp[$\ast$]{Corresponding author. \href{saadb@chbe.gatech.edu}{saadb@chbe.gatech.edu}}

\received{Date}{0}{Year}
\revised{Date}{0}{Year}
\accepted{Date}{0}{Year}



\abstract
{Many organisms utilize group aggregation as a method for survival. The freshwater oligochaete, \textit{Lumbriculus variegatus} (California blackworms) form tightly entangled structures, or worm ``blobs", that have adapted to survive in extremely low levels of dissolved oxygen (DO). Individual blackworms adapt to hypoxic environments through respiration via their mucous body wall and posterior ciliated hindgut, which they wave above them. However, the change in collective behavior at different levels of DO is not known. Using a closed-loop respirometer with flow, we discover that the relative tail reaching activity flux in low DO is $\sim$75x higher than in the high DO condition. Additionally, when flow rate is increased to suspend the worm blobs upward, we find that the average exposed surface area of a blob in low DO is $\sim$1.4x higher than in high DO. Furthermore, we observe emergent properties that arise when a worm blob is exposed to extreme DO levels. We demonstrate that internal mechanical stress is generated when worm blobs are exposed to high DO levels, allowing them to be physically lifted off from the bottom of a conical container using a serrated endpiece. Our results demonstrate how both collective behavior and the emergent generation of internal mechanical stress in worm blobs change to accommodate differing levels of oxygen. From an engineering perspective, this could be used to model and simulate swarm robots, self-assembly structures, or soft material entanglements.}


\maketitle
\section{Introduction}
\begin{figure*}[h!t]
\centering
\includegraphics[width=\hsize]{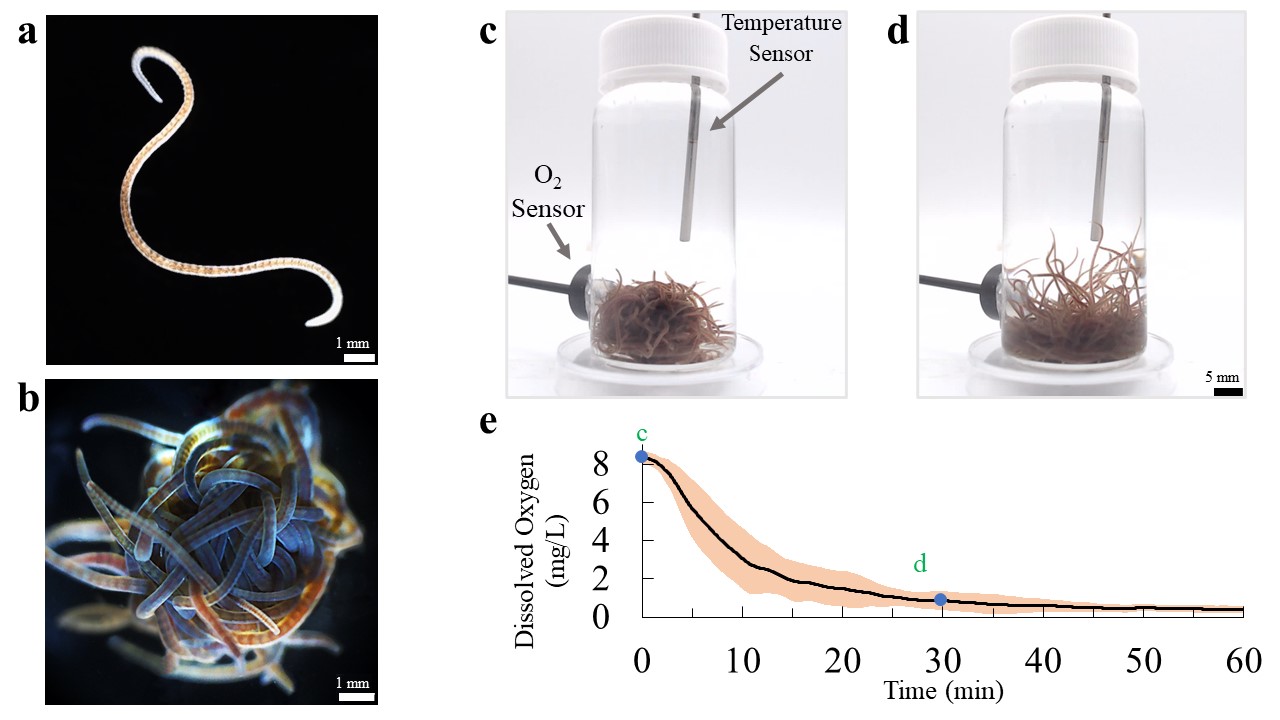}
\caption[Behavioral difference of California blackworms in two oxygen environments]{\textbf{Behavioral difference of California blackworms in two oxygen environments.} \textbf{(a)} An individual \textit{L. variegatus} (California blackworm). \textbf{(b)} A worm blob consisting of 12 individual blackworms. \textbf{(c)} 1 g of blackworms in a 20 mL glass vial at high dissolved oxygen (DO) level (8.3±0.3 mg/L). High DO levels induce stress among blackworms and they create a spherical-shaped structure. The oxygen sensor is affixed on the left side of the container and the temperature sensor is inserted through the cap from above. The vial is filled with distilled water. \textbf{(d)} In 30 minutes, DO reaches very low levels of 0.9±0.5 mg/L. At this point, tail reaching behavior is pervasive to supplement respiration. \textbf{(e)} Average dissolved oxygen as a function of time. Shaded region corresponds to standard deviation with n=7 individual trials. 
\textbf{(Supplementary Movie S1)}}
\label{fig:respiration}
\end{figure*}

Living or active matter collectives are composed of many entities that utilize energy for activity (\citealt{2020Das}). In nature, these collectives consist of organisms of diverse sizes, from single-cell organisms like bacteria, to insects and birds, to enormous animals such as whales, to reproduce, migrate, or survive (\citealt{1999Parrish, 2012Vicsek, 2019Smith}). While many of these collectives consist of discretely separated individuals such as a bird in a flock, there are groups that form a physically entangled structure such as \textit{Caenorhabditis elegans} (\citealt{2019Ding, 2015Artyukhin}), earthworms (\citealt{2022Patnaik, 2012Zirbes}), honeybees (\citealt{2018Peleg}), fire ants (\citealt{2021Wagner, 2016Hu}), daddy longlegs (\citealt{2021Wagner}), and the freshwater-based oligochaete \textit{Lumbriculus variegatus} (California blackworms) (Fig.\ref{fig:respiration}a). These worms can form a highly compact and dense worm ``blob" (Fig.\ref{fig:respiration}b) structure that exhibits properties comparable to a highly viscous non-Newtonian fluid when outside of granular substrate or detritus (\citealt{2020aDeblais, 2020bDeblais, 2021Nguyen, 2021Ozkan-Aydin}). As with most oligochaetes, this agglomeration is due to their thigmotactic behavior, which causes them to wrap around and tangle with each other for survival against desiccation, predation, and other factors (\citealt{2015Timm, 2010Zirbes}). In this globular structure, blackworms are also capable of collectively and emergently locomoting away from excessive light or warm temperatures due to their negative phototactic or thermotactic behaviors (\citealt{2015Timm, 1990Drewes, 2021Ozkan-Aydin}). 

Many oligochaetes are benthic organisms and live at the bottom of different bodies of waters where several parameters, such as temperature, oxygenation, or substrate-type vary widely. Other oligochaetes live in oxygen-depleted sulfur caves where DO is sustained at 0.5 mg/L due to high levels of hydrogen sulfide (\citealt{2016Fend,2017Giere}. Blackworms are one of many invertebrates that have adapted to use little oxygen, which may arise from changes within the environment, including shallow water zones or from foliage covering the surface of the water such as duckweed or algae (\citealt{2008Vymazal, 1998Hamburger, 2018Leys, 2019Ceschin}). Individual blackworms have adapted to these low dissolved oxygen (DO) conditions by respiring through their mucosal body wall and by having the ability to supplement respiration through their posterior ciliated hindgut (\citealt{2015Timm, 2021Glasby}). By doing so, individuals can survive long periods of time ($>$10 days) in DO conditions down to 0.7 mg/L (\citealt{2008Mattson}). In granular substrates, these worms typically burrow and anchor their heads while simultaneously lifting and waving their tails to circulate surrounding water for oxygen (\citealt{1990Drewes, 2015Timm}). Additionally, because blackworms can also reduce their pulsation rates, which reduce their metabolic rates (\citealt{2000Penttinen, 1998Hamburger, 1999Lesiuk}). However, one factor that has not been investigated previously is how a large collective of aerobically-respiring and physically-entangled worms, which by themselves create a localized zone of low DO, are capable of surviving together in these conditions. 

Therefore, in this study, we assess the emergent behavioral change of worm blobs in normoxic (which we define as DO levels $>$8.0 mg/L) and hypoxic ($<$2.0 mg/L) conditions. Specifically, we ask the questions: (i) how does a worm blob collectively behave under high and low DO conditions? and (ii) does the internal mechanical stress generated through entanglements in the blob vary between these two DO conditions, possibly for collective survival benefits?

\section{Materials and methods}\label{sec2}
\subsection{\textbf{Animals}}\label{subsec1}
We procure California blackworms (length 30.2±7.4 mm, diameter 0.6±0.1 mm, mass 7.0±2.4 mg) from Aquatic Foods $\&$ Blackworms Co. and from Ward's Science. They are reared in a plastic storage box (35 X 20 X 12 cm) with filtered water and stored in a refrigerator maintained at 4$^{\circ}$C (\citealt{2021Martinez}). We feed blackworms tropical fish flakes once a week and swap their water every day. Prior to experiments, worms are habituated in room temperature water ($\sim$21$^{\circ}$C) and room lighting for at least 1 hr. Studies with blackworms do not require approval by an institutional animal care committee. For each experiment, we measure 1±0.01 g ($\sim$150 individuals) of worms by removing as much water as possible using a pipette and absorbents. 

\subsection{\textbf{Data Acquisition}}\label{subsec3}
A Logitech$^\copyright$ C920x webcam is used to record all experiments in a photobox with fixed lighting ($\sim$500 lux). Temperature-compensated oxygen measurements are recorded using Pyroscience FireSting PRO coupled with a contactless sensor spot and a temperature probe. The oxygen sensor is calibrated using oxygen scavenger capsules ($\geq$ 98\% sodium sulphite) for the lower threshold and using aerated water for the upper threshold. The temperature of the water for each experiment is maintained at room temperature, which varied from 21.0±0.3$^{\circ}$C.

\subsection{\textbf{Data Analysis}}\label{subsec4}
For each experiment, recordings are processed and converted into AVI format using Adobe$^\copyright$ Premiere Pro. The AVI is then imported into ImageJ for image analysis. After background subtraction, each stack is binarized using the same threshold settings. 
For the tail reaching activity experiment, a region of analysis is set to a height $2/3$ the length of an average worm ($\sim$22.5 mm) from the bottom of the tube to calculate an average pixel value. We select this height to allow sufficient data visualization without excessive data overlap, which occurs when the height is set too low or too high and to allow worms to physically reach upwards. A line of one pixel in thickness is drawn across the inner diameter of the tube and the average pixel intensities are calculated in ImageJ. We refer to this region as the tail reaching ``relative activity flux" (Fig.\ref{fig:activity}b,c).
For the floating worm blob experiments, we assume that the dark pixels of the binarized image of the worm blob provide an approximation of the projected exposed surface area to compare between the two DO environments. The total area of each stack is computed using the Analyze Particles tool in ImageJ. 
For the lifting experiments, we calculate the center of mass height ($H_{COM}$) from each binarized image using the Measure tool in ImageJ. We note that the center of mass technically represents the center of geometry, as we use this metric as a representation for the blob mass.

\section{Results}\label{sec4}
Over the span of an hour in a 20 mL glass vial, we observe that DO shifts to hypoxic levels, which we define as $<$2 mg/L (Fig.\ref{fig:respiration}e). Concurrently, this observed shift in DO results in a dramatic change in worm blob behavior (Fig.\ref{fig:respiration}c,d), from a tightly spherical-shaped and immobile structure to a highly active and disentangled collective. This shift in behavior is robust and is highly repeatable (n=7 individual trials).

\subsection{\textbf{Closed-loop Respirometer with Flow}}\label{subsec5}

\begin{figure}[h!]
\centering
\includegraphics[width=\hsize]{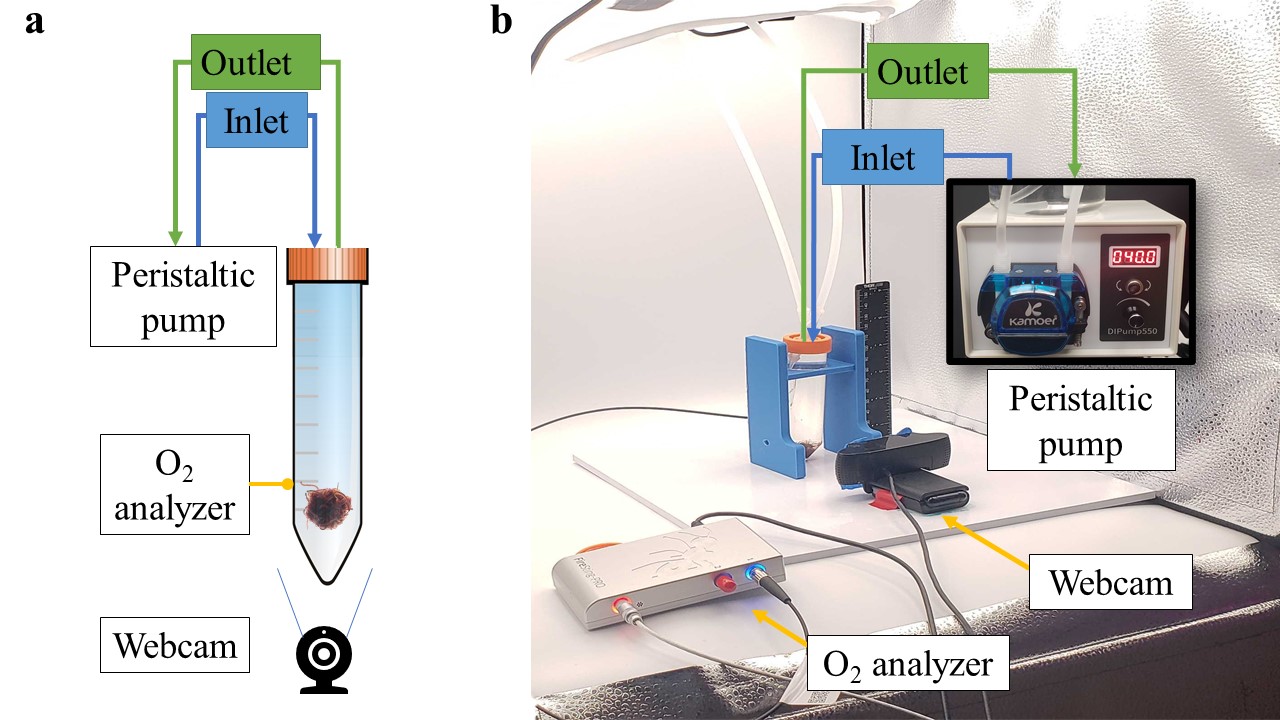}
\caption[Closed-loop respirometer with flow setup.]{\textbf{Closed-loop respirometer setup.}  \textbf{(a)} Schematic of the setup.  \textbf{(b)} Image of setup showing a 50 mL polypropylene centrifuge tube that is used as the experimental chamber. Two 1/4” water tubes are inserted through the plastic cap and sealed with superglue. A 200 µL pipette tip is fixed onto the end of the inlet tube to direct flow. Tubes are connected to a peristaltic pump to provide fluid flow. A contactless oxygen sensor spot is placed on the tube wall, 35 mm from the bottom. A webcam records all experiments at 30 FPS.}
\label{fig:flowsetup}
\end{figure}

\begin{figure}[ht!]
\centering
\includegraphics[width=\hsize]{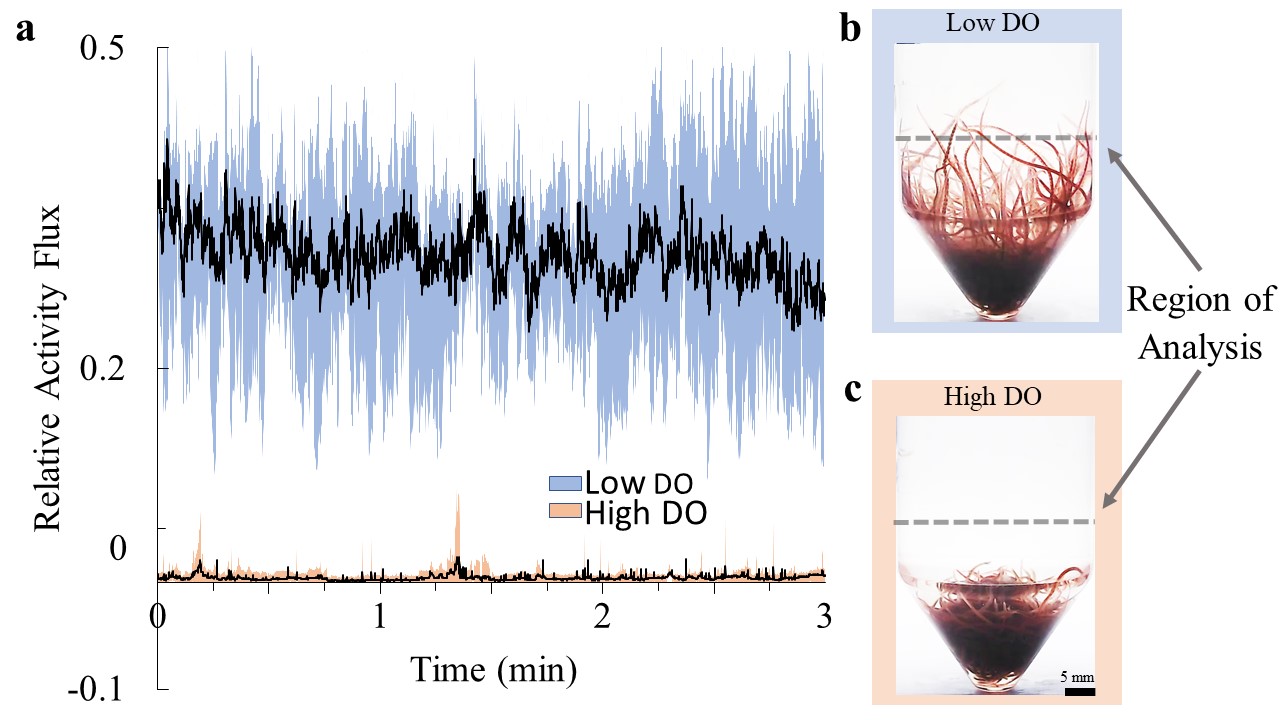}
\caption[Tail reaching activity of California blackworms]{\textbf{Tail reaching activity of California blackworms in two oxygen environments} Measuring relative activity levels of worm blobs (1 g) as a function of time in two dissolved oxygen environments. A closed-loop chamber with peristaltic pump-driven water flow is used to circulate water at a rate of $0.23 \pm0.1$ mL/s. Relative activity (mean gray value) is measured at the “Region of Analysis”, which is set $2/3$ the average length of the worms ($\sim$22.5 mm) from the bottom of the conical tube. \textbf{(a)} shows the relative activity level as a function of time in both environments. Data represents the averages and the shaded regions correspond to span for n=5 trials. Each trial is recorded for 4 mins and the last 3 mins are extracted for image analysis. \textbf{(b)} In high DO ($>8.5$ mg/L), blackworms form tight entanglements with little to no tail reaching activity. In low DO $<2$ mg/L \textbf{(c)}, blackworms lift their tails to supplement respiration which increases activity. \textbf{(Supplementary Movie S2).}}
\label{fig:activity}
\end{figure}

The purpose of the first experiment is to quantify tail reaching activity, which was described earlier as a behavior to supplement respiration in low DO conditions. We create a closed-loop respirometer using a clear conical tube with a modified cap connected to a peristaltic pump (Fig.\ref{fig:flowsetup}a,b). Here, the pump is added to provide some flow to mix and to remove dissolved oxygen gradients within the system. 
Two 1/4" holes are drilled on the top of the cap to allow the insertion of water tubes, which are sealed with superglue. A 200 µL pipette tip is fixed onto the inlet tube to direct water flow to the wall and around the bottom of the conical tube. The oxygen sensor spot is attached on the inside wall 35 mm from the bottom of the tube using silicone glue. Finally, the container is pressure tested for leaks and rinsed with distilled water using the peristaltic pump for 1 hour. 

To create a hypoxic or low DO environment, 1 g of worms are placed into the enclosed tube and water is recirculated at 1 mL/s for 2 hrs. After reaching DO levels of $<$2.0 mg/L, the peristaltic pump flow rate is lowered to 0.2 mL/s to provide low flow for recirculation so as to not disturb the worms. To create a normoxic environment, which we define as high DO levels $>$8.0 mg/L, the outlet tube from the pump is disconnected and placed into an aerated water reservoir. A separate tube is installed on the pump to intake this water. Care is taken to prevent the intake of air bubbles which are removed if present in the system. 
The behavior is then recorded for 4 minutes where the last 3 minutes are extracted for data analysis. Each experiment is repeated 5 times in their respective DO environments. 

We quantify the change in behavior in the stagnant environment shown in Fig.\ref{fig:respiration}c,d via image analysis. Fig.\ref{fig:activity}b,c shows similar behaviors compared to the stagnant water environment. With respect to our defined conditions for quantifying activity, the average relative activity flux is 30.6±8.5$\%$ and 0.4±0.6$\%$ for low and high DO conditions, respectively (Fig.\ref{fig:activity}a). Additionally, the average respiration rate in the low and high DO conditions are 0.8±0.2 mg/L$\cdot$hr and 3.1±0.4 mg/L$\cdot$hr.

\begin{figure}[!ht]
\centering
\includegraphics[width=\hsize]{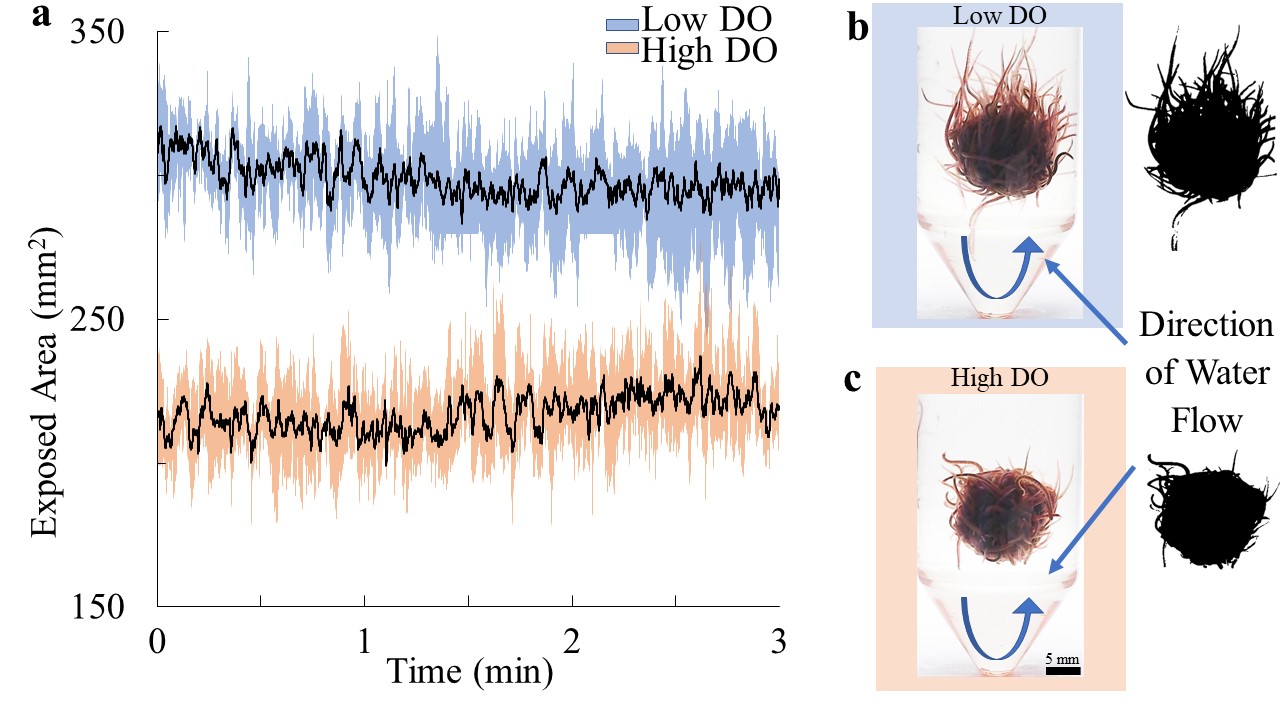}
\caption[Floating worm blob]{\textbf{Floating worm blobs in two oxygen environments} Measuring exposed surface area of worm blobs (1 g) as a function of time in two dissolved oxygen environments. \textbf{(a)} shows the exposed surface area as a function of time in both environments. Shaded region corresponds to span with n=5 trials. A closed-loop chamber with peristaltic pump-driven water flow at a rate of 0.6±0.1 mL/s is used to levitate worm blobs and observe exposed surface area. \textbf{(b)} In high DO ($>$8.0 mg/L), they form a tight, spherical ball. In low DO $<$2.0 mg/L (c), blackworms disentangle their tails to supplement respiration which increases exposed surface area. Exposed area is extracted using image thresholding techniques in ImageJ, as shown on the right of each corresponding image. \textbf{(Supplementary Movie S3).}}
\label{fig:float}
\end{figure}
The second experiment rules out interference with the conical shape of the tube and provides equal exposure of DO to the worm blobs. Due to the direction of the pipette tip, increasing fluid flow directs water towards the bottom of the worm blob which lifts it upwards, allowing us to see the entire structure and to quantify a relative projected exposed surface area. Continuing from the oxygen conditions of the previous experiment, the peristaltic pump flow rate is increased to 3x from the first experiment (0.6 mL/s) which is predetermined to provide enough lift to keep the structure constantly afloat. The behavior is then recorded for 4 minutes where the last 3 minutes are extracted for data analysis. Each experiment is repeated 5 times in their respective DO environments. 

Using our assumptions to calculate the approximate exposed surface area, the average exposed surface is 298.9±16.3 mm$^2$ and 217.1±13.6 mm$^2$ in low and high DO conditions, respectively (Fig.\ref{fig:float}a). The average respiration rate in the low and high DO are 1.4±0.2 mg/L$\cdot$hr and 15.6±4.8 mg/L$\cdot$hr, respectively, which are higher than the previous experiment. We believe this is from removing wall effects and increasing the flow rate, both of which increase the worm blob's exposure to DO. This also shifts access to DO from mostly diffusive to forced convection, modulating and increasing gas exchange within a tightly entangled structure (\citealt{1989Patterson, 2002Bird}). 

In both experiments, worms in low DO (Fig.\ref{fig:float}b) appear to show less packing compared to the near spherical shape of the worms in high DO (Fig.\ref{fig:float}c) as worms disentangle and extend their tails to supplement respiration. 

\subsection{\textbf{Blob-on-a-Stick: Lifting Worm Blobs}}\label{subsec6}

\begin{figure}[ht!]
\centering
\includegraphics[width=\hsize]{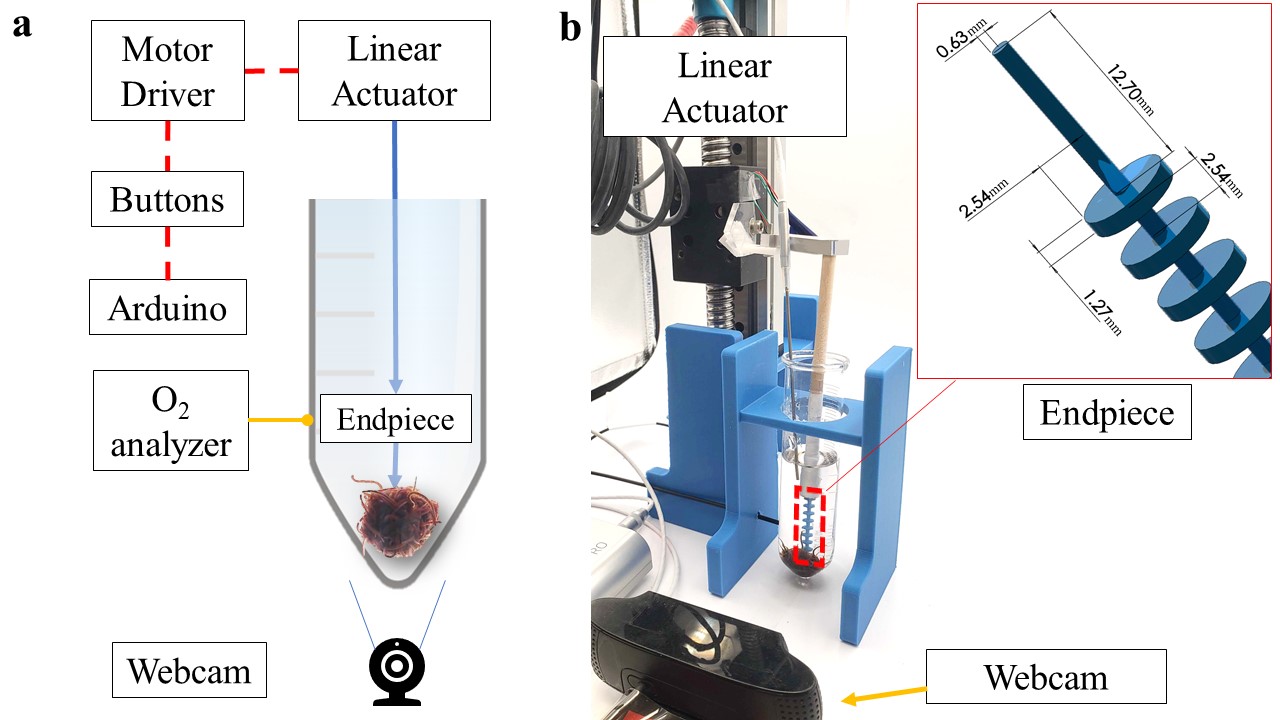}
\caption[Lifting setup.]{\textbf{Lifting setup.} A linear actuator programmed through an Arduino is used to provide a constant upward velocity of 0.5±0.01 mm/s and downward velocity of 2.0±0.01 mm/s actuated by buttons. A 3D-printed serrated endpiece is attached to the linear actuator using a small screw and a wooden dowel. A contactless oxygen sensor spot is placed on the tube wall, 20 mm from the bottom. A webcam records all experiments at 1 FPS using MATLAB image acquisition.}
\label{fig:liftsetup}
\end{figure}

We assume that the amount of packing from previous experiments is related to the internal mechanical stresses generated from the contraction of blackworms, which depends on DO conditions. To quantify this, we physically lift worm blobs using a 3D-printed serrated endpiece driven by a linear actuator and track its center of mass height ($H_{COM}$) via image analysis. 

After creating the hypoxic environment using the closed-loop respirometer setup as described above, the cap is removed and the linear actuator assembly is installed. The schematic is shown on Fig.\ref{fig:liftsetup}a and the experimental setup is shown in Fig.\ref{fig:liftsetup}b. The linear actuator is manually controlled to provide constant upward or downward movement via an Arduino. The upward movement occurs at a constant velocity of 0.5±0.01 mm/s. The downward movement occurs at a constant velocity of 2.0±0.01 mm/s and is only used to reset the position of the setup. A serrated endpiece is 3D-printed and is used to provide a sufficient gripping surface for the worms. This endpiece is attached to the linear actuator by a wooden dowel. Once the endpiece is lowered to the bottom of the container, the worms are carefully mixed using a pipette without adding air bubbles. The worms are then given 5 minutes to habituate around the endpiece. Soon after, a 3-minute recording at 1 FPS is started and the up button is pushed for 30 seconds. During image processing, 90 seconds of recording (from pushing the button) are extracted for analysis. For high DO experiments, an aerator is used to raise DO levels. 

\begin{figure*}[!ht]
\centering
\includegraphics[width=\hsize]{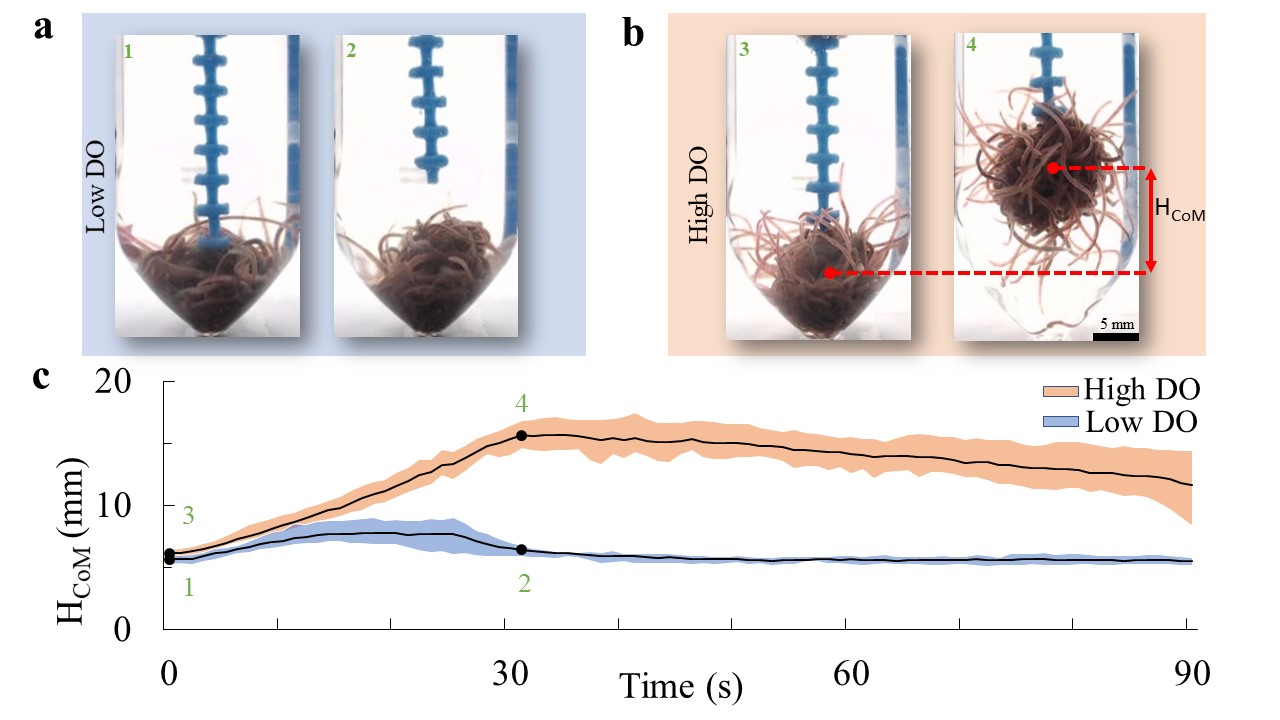}
\caption[Blob-on-a-stick]{\textbf{Blob-on-a-stick: Assessing internal mechanical stress} Measuring position of center of mass height (H$_{CoM}$) of worm blob structure as a function of time in two dissolved oxygen environments. Using a 3D-printed serrated endpiece, worm blobs are lifted up by 15 mm using a linear actuator. (a) In low DO ($<$2.0 mg/L), they display less internal mechanical stress by entangling loosely with one another. In high DO ($>$8.0 mg/L) (b), they show increased amounts of internal mechanical stress, enough to be lifted using the endpiece. (c) Position of the H$_{CoM}$ as a function of time in both environments. Shaded region corresponds to the span of $n=5$ trials. 
\textbf{(Supplementary Movie S4).}}
\label{fig:blob}
\end{figure*}

In contrast to low DO conditions (Fig.\ref{fig:blob}a), we are able to lift the worm blob from the bottom of the conical tube in high DO conditions (Fig.\ref{fig:blob}b), remaining suspended for the duration of each trial. The worm blob's $H_{COM}$ in high DO reached a maximum of 12.7±2.9 mm as compared to 6.2±0.9 mm in low DO (Fig.\ref{fig:blob}a).

\section{Discussion}\label{sec5}
Blackworms can adjust their physiology to survive in many environmental conditions as confirmed from several studies conducted at the individual worm level (\citealt{2000Penttinen,1998Hamburger, 1999Lesiuk}). However, at the collective level, localized low DO areas can form when a large number of aerobically-respiring worms entangle together. Physiologically, our results confirm that blackworms can adjust their respiration rates to adapt to varying DO environments. Furthermore, blackworms collectively adapt by lifting their tails upwards, which allows individuals to reach higher DO levels, away from the collective. Consequently, this also allows them to mix DO from above which may benefit shorter conspecifics. From the metrics we set to measure activity flux and exposed surface area, our results indicate an average increase of $\sim$75x in relative activity flux and a $\sim$1.4x increase in exposed surface area when comparing high DO to the low DO conditions. In the opposite sense, high DO may signal worms to strongly entangle with one another, as shown in our lifting experiment where the internal mechanical stress is high enough to allow us to lift the worms using a serrated endpiece. We believe that high DO conditions could arise from torrential conditions, such as in high winds where strong currents and mixing occur or from rainstorms, where fresh oxygenated water distributed throughout their habitat. In both situations, individual worms could be displaced away from substrates, decreasing their chance of survival. One example is how sessile \textit{Gorgasia sillneri} (garden eels) change their posture depending on the current. In higher currents, individual eels shorten their exposed surface area to reduce drag force as they forage for microorganisms (\citealt{2022Ishikawa, 2018Khrizman}). Similarly, individuals from \textit{Apis mellifera} (European honeybees) tree-hanging colonies respond to windy conditions by spreading out and by flattening the cluster to maintain structural stability (\citealt{2018Peleg}). Overall, it would be beneficial for blackworms to strongly entangle with conspecifics or other objects to increase their chance of survival. 
Our work could inspire engineering design of amorphously entangled robotic matter such as shape-holding active matter collectives, or self-organizing systems (\citealt{2019Savoie, 2021Ozkan-Aydin, 2021Chvykov}).

\subsection{\textbf{Limitations}}\label{subsec7}
Every aerobically-respiring organism will eventually die when exposed to anoxic oxygen levels for long periods of time due to the depletion of glycogen storage. For blackworms, this would imply that their relative activity flux will eventually decrease to zero. Previous studies by \citeauthor{2008Mattson} determined that 100$\%$ of blackworms survived in a 10-day exposure to low DO levels down to 0.7 mg/L. Additionally, a different study by \citeauthor{1998Hamburger} reported that the 50$\%$ survivorship of blackworms in lower DO levels of $\sim$0.15 mg/L was 40 to 50 days. Both studies are outside the timescale of our experiments. 
Furthermore, to limit the scope of our work, we did not vary the population of blackworms which was fixed to a mass of 1 g. Also, we did not test their behavior in their natural environment due to difficulty in controlling DO.

\section{Conclusions}\label{sec6}
In this study, we assessed the emergent behavioral change of worm blobs in two extreme oxygen conditions. In low DO, worm blobs increased their tail reaching activity to acquire dissolved oxygen away from the population. Furthermore, worm blobs inherently increased their exposed surface area when levitated upwards from the bottom of the container and provided with equal exposure to DO. In higher DO conditions, blackworms entangled strongly with one another, increasing internal mechanical stresses enough for us to lift the collective with a 3D-printed serrated endpiece. We believe blackworms tangle strongly with one another to prevent being displaced from heavy weather conditions, signalled by high DO conditions.

\section{Supplementary data}\label{sec9}
Supplementary data available at \textit{ICB} online.

\section{Competing interests}
There is NO Competing Interest.

\section{Author contributions statement}

H.T. and E.K. designed and carried out the experiments, analyzed, and interpreted the results. D.I.G. and M.S.B. reviewed the design and execution of experiments, the data analysis, the interpretations, and the manuscript. All authors contributed to writing the manuscript.

\section{Acknowledgments}
H.T. acknowledges funding support from the NSF graduate research fellowship program (GRFP) and Georgia Tech's President's Fellowship. E.K. acknowledges funding support from Georgia Tech’s Presidential Undergraduate Research Award (PURA). D.I.G. acknowledges funding support from ARO MURI award (W911NF-19-1-023) and NSF Physics of Living Systems Grant (PHY-1205878). M.S.B. acknowledges funding support from NIH Grant R35GM142588 and NSF Grants CAREER 1941933 and 1817334.
We thank members of the Bhamla lab and Goldman lab for useful discussions.We especially want to thank Dr. Christopher Pierce for the image analysis training and Dr. Philip Matthews for providing expert feedback related to topics of this work.
 
\bibliographystyle{apalike}
\bibliography{reference}


\end{document}